\documentclass[12pt]{iopart}
\usepackage[bookmarks,dvips,pdfhighlight=/O,pdfstartview=FitH]{hyperref}
\usepackage{iopams}
\usepackage{graphicx,amsopn,color}

\DeclareMathOperator{\im}{Im}

\begin{document}

\title{Interpretation of Recent SPS Dilepton Data}

\author{Hendrik van Hees and Ralf Rapp}

\address{Cyclotron Institute and Physics Department, Texas A\&M
  University, College Station, Texas 77843-3366, USA}

\ead{hees@comp.tamu.edu}

\begin{abstract}
  We summarize our current theoretical understanding of in-medium
  properties of the electromagnetic current correlator in view of recent
  dimuon data from the NA60 experiment in In(158~AGeV)-In collisions at
  the CERN-SPS. We discuss the sensitivity of the results to space-time
  evolution models for the hot and dense partonic and hadronic medium
  created in relativistic heavy-ion collisions and the contributions
  from different sources to the dilepton-excess spectra.
\end{abstract}

\section{Introduction}

A main goal of relativistic heavy-ion collisions is to understand the
phase structure of strongly interacting matter as inferred from Quantum
Chromodynamics (QCD). Lattice-QCD calculations at high temperatures find
a transition from hadronic matter to a deconfined and chirally restored
quark-gluon plasma (QGP), where strong modifications of the spectral
properties of hadrons are expected.  In the low-mass region
($M$$\lesssim$1~GeV), dileptons are valuable probes for medium
modifications of the low-mass vector mesons, $\rho$, $\omega$ and
$\phi$. In particular the $\rho$ meson, which has a small lifetime
($\sim$1.3~fm/$c$) already in the vacuum, mostly decays inside the hot
and dense medium. Since dileptons do not suffer strong final-state
interactions, they provide direct information about the in-medium
spectral properties of the light vector mesons. In Pb(158~AGeV)-Au
collisions at the Super-Proton-Synchrotron (SPS), an enhancement of
dielectrons at low invariant masses~\cite{Agakichiev:2005ai} indeed
indicates strong medium modifications of the electromagnetic (e.m.)
current correlator. Recently, with the better mass resolution and
statistics of the NA60 dimuon spectra in In(158~AGeV)-In
collisions~\cite{na60-06} a discrimination of models assuming different
mechanisms underlying the dilepton excess has become possible: Models
predicting a large in-medium broadening of the $\rho$-meson spectrum
with little mass shifts are favored by the data over those implementing
a dropping $\rho$ mass~\cite{HR06a,HR06b}.

In addition to a careful description of the e.m. current correlator in
thermal partonic and hadronic matter, a reliable theoretical
interpretation of the measured dilepton spectra requires two more
ingredients: (i) a model for the evolution of the hot and dense medium,
typically including a QGP at the early collision stages, followed by a
transition to a hadronic phase and subsequent expansion until thermal
freeze-out; (ii) consideration of non-thermal dilepton sources like
primordial quark annihilation (Drell-Yan), decays of $\rho$ mesons which
have not thermally equilibrated with the medium (``Corona effect'') and
$\rho$ decays after thermal freeze-out.

In the following we briefly summarize our model to describe emission of
dileptons from a thermalized strongly interacting medium
(Sec.~\ref{sec.inmedvecmes}) and refine our previous
description~\cite{HR06a} of non-thermal sources (Sec.~\ref{sec.dilept}).

\section{Dilepton emission from a thermal source}
\label{sec.inmedvecmes}
Dilepton emission from a thermal source of strongly interacting
particles is given by~\cite{MT84}
\begin{equation}
\label{1}
\frac{\rmd N}{\rmd^4 x \rmd^4 q}=-\frac{\alpha^2}{\pi^3}
\frac{L(M^2)}{M^2} \im \Pi_{\mathrm{em}}^{(\mathrm{ret})}(M,q) f_B(q_0),
\end{equation}
where $\alpha$$\simeq$1/137, $f_B$ is the Bose distribution function and
$L$ a lepton-phase space factor. $\Pi_{\mathrm{em}}^{(\mathrm{ret})}$
denotes the retarded in-medium e.m. current-correlation function, which
in the vacuum, and for $M$$<$1~GeV, is saturated by the vector-meson
spectral functions:
\begin{equation}
\label{2}
\im \Pi_{\mathrm{em}}^{(\mathrm{ret})}=\sum_{V=\rho,\omega,\phi}
\frac{m_V^4}{g_V^2} \im D_V.
\end{equation}
In a hot and dense hadronic medium, the current correlator is evaluated
using many-body theory based on an effective hadronic Lagrangian.  The
model parameters (masses, coupling constants and form factors) are
adjusted to empirical hadronic decay rates, scattering data like $\pi N
\rightarrow V N$ and photo absorption on nucleons and nuclei. The
vector-meson propagators are evaluated at finite temperature and
density,
\begin{equation}
D_V(M,q)=(M^2-m_V^2-\Sigma_{VP}-\Sigma_{VM}-\Sigma_{VB})^{-1},
\end{equation}
where the three self-energy contributions are due to the interactions of
the vector mesons with their pseudoscalar-meson cloud as well as mesons
and baryons in the medium~\cite{Rapp:1999us}.  For the $\rho$ meson,
hadronic many-body models predict a strong broadening but small mass
shifts with increasing temperature and density. For the $\omega$ meson,
one also finds broadening ($\Gamma_{\omega}^{\mathrm{med}} \simeq
50\;\mathrm{MeV}$ at nuclear-matter density,
$\varrho_0=0.16/\mathrm{fm}^3$)~\cite{Rapp01}. The $\phi$ meson is
expected to broaden in hot and dense matter mostly due to modifications
of its kaon cloud. Data on nuclear photoproduction of $\phi$ mesons give
absorption cross sections leading to an in-medium width of about 50~MeV
at $\varrho_0$~\cite{CROTV04}. In the intermediate-mass region ($M >
1$~GeV), the e.m. correlator is dominated by four-pion type states. We
describe the pertinent dilepton emission in terms of the empirically
known vacuum e.m. correlator augmented by model-independent effects of
chiral mixing to leading order in $T$~\cite{DEI90,HR06a}. In the QGP
thermal dilepton radiation is due to $q\bar{q}$ annihilation for which
we employ the hard-thermal loop improved perturbative QCD
result~\cite{bpy90}.

\section{Dilepton spectra from the medium created in heavy-ion collisions}
\label{sec.dilept}
\begin{figure}[!t]
\begin{center}
\begin{minipage}{0.32 \textwidth}
\includegraphics[width=\textwidth]{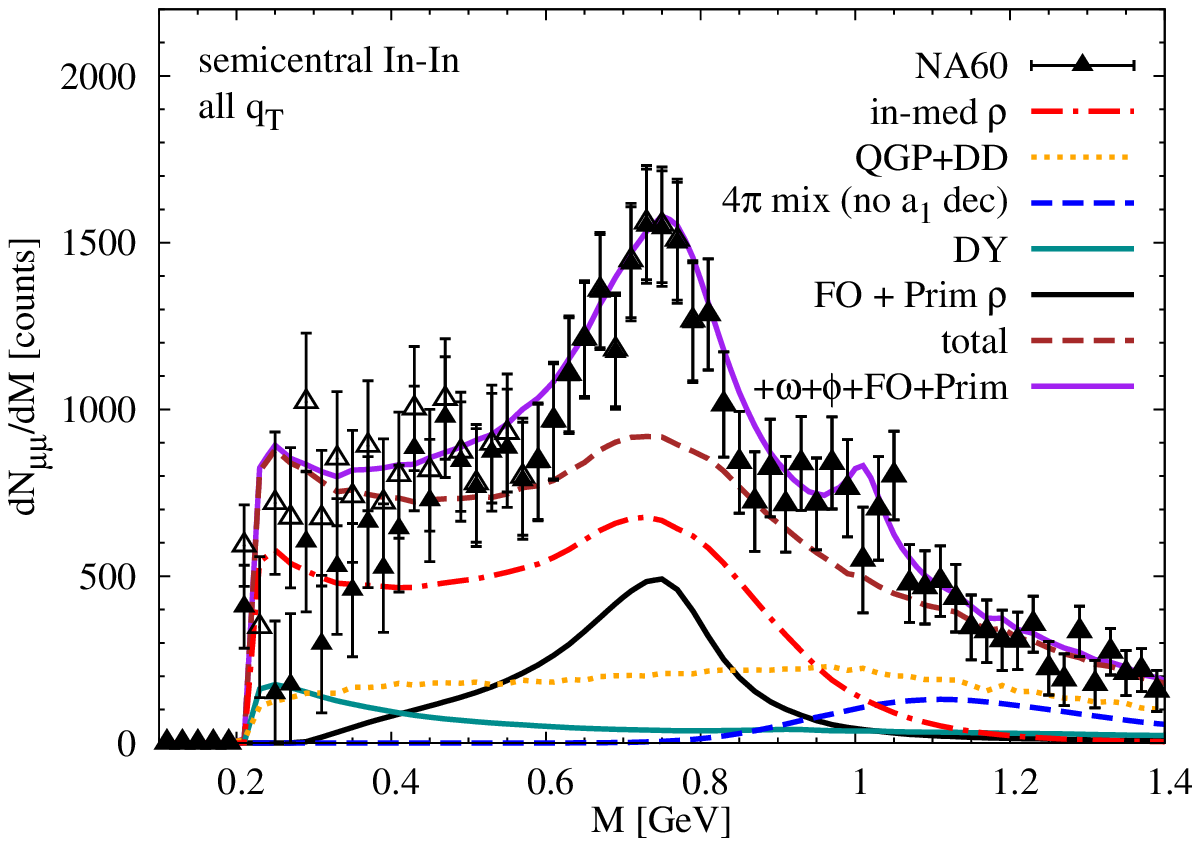}
\end{minipage}
\begin{minipage}{0.32 \textwidth}
\includegraphics[width=\textwidth]{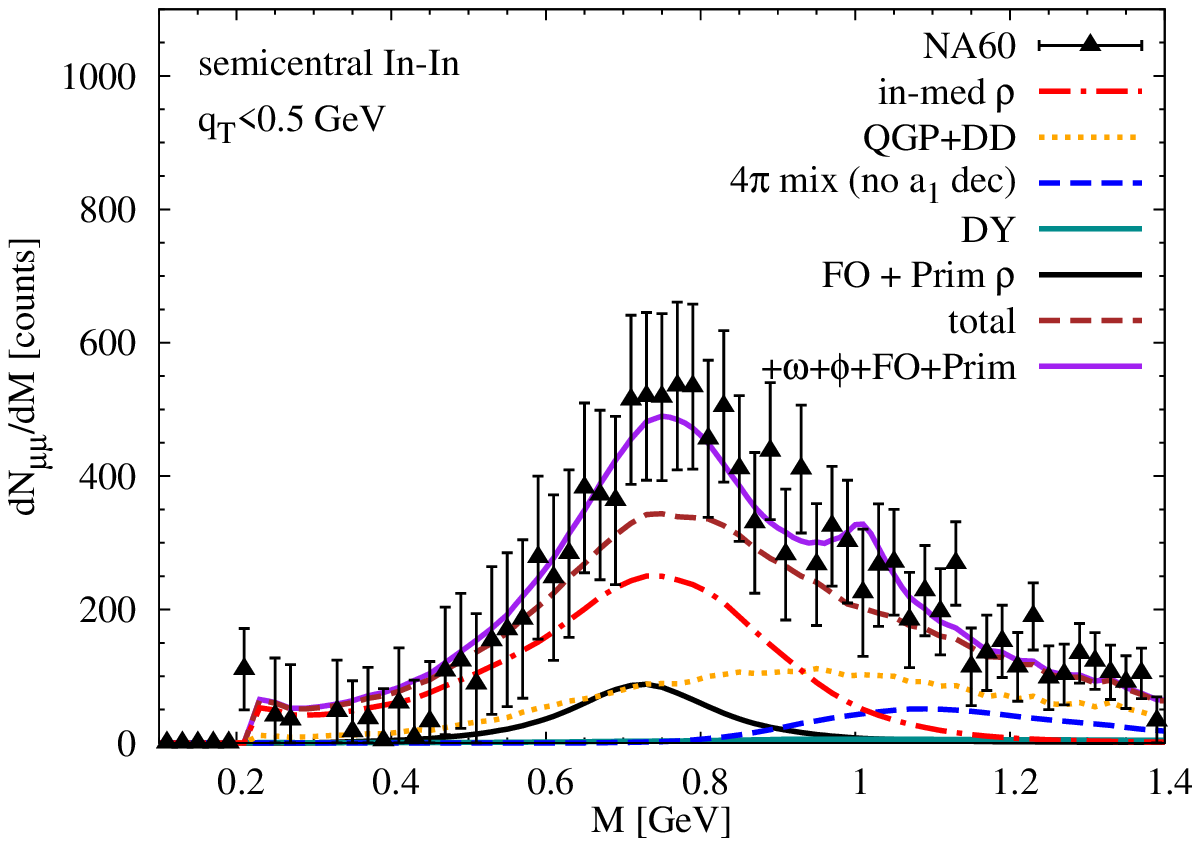}
\end{minipage}
\begin{minipage}{0.32 \textwidth}
\includegraphics[width=\textwidth]{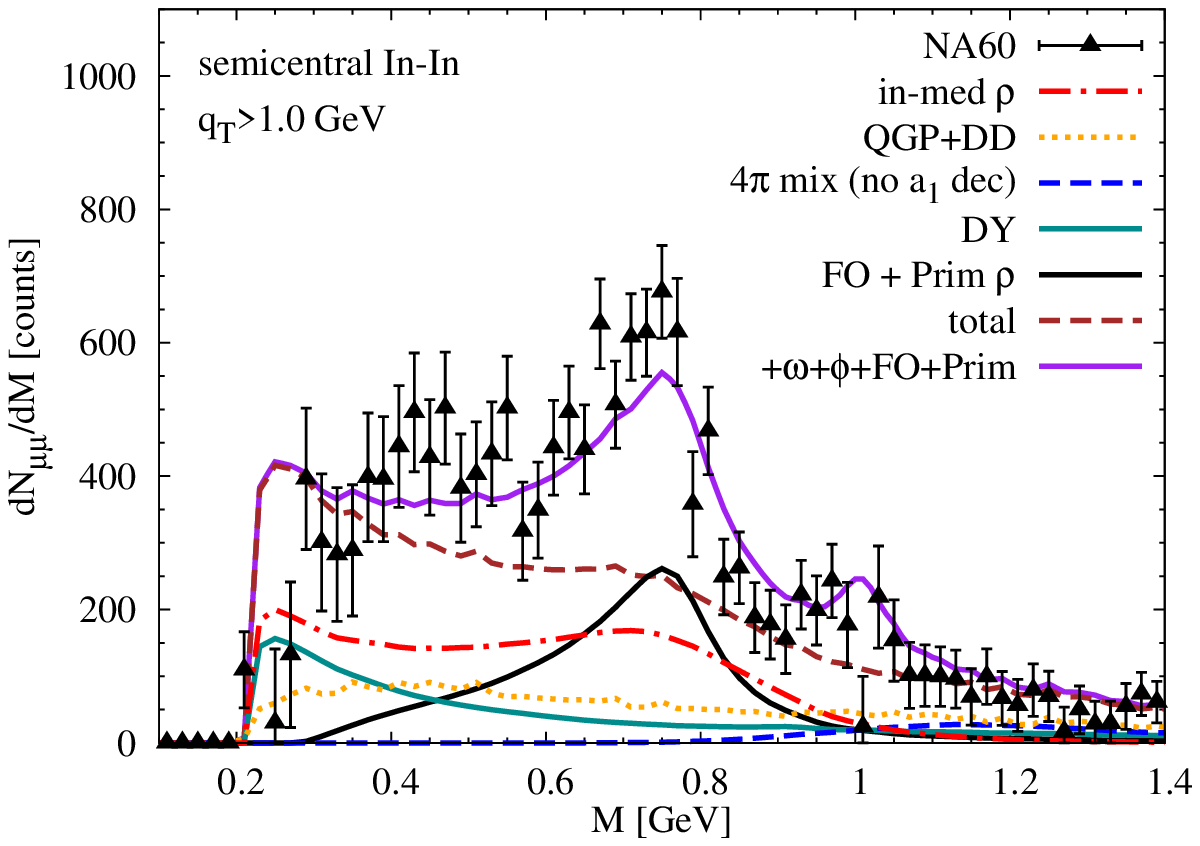}
\end{minipage}
\end{center}
\vspace*{-0.6cm}
\caption{Excess dimuon spectra in semicentral In(158~AGeV)-In collisions
  in different $q_T$ regions (from left to right: all $q_T$,
  $q_T<0.5$~GeV, $q_T>1$~GeV) compared to emission from QGP, in-medium
  $\rho$, $\omega$ and $\phi$ mesons, four-pion annihilation, correlated
  $D\bar{D}$ decays, primordial Drell-Yan annihilation and decays of
  $\rho$ mesons after thermal freeze-out and non-equilibrated primordial
  $\rho$'s. \label{Fig.1}}
\vspace{-2mm}
\end{figure}
We evaluate the dilepton spectrum from Eq.~(\ref{1}) by integration over
the space-time history of the collision based on a cylindrical thermal
fireball expansion~\cite{Rapp:1999us,HR06a} with a linear transverse
flow profile, $v_\perp(r,t)$=$v_s(t) r/R(t)$ where $v_s(t)$=$a_\perp t$
and $a_\perp$=$0.085c^2$/fm as an upper estimate of flow properties from
hydrodynamic models~\cite{Kolb:2003dz} ($R(t)$: fireball radius). For
central In(158~AGeV)-In collisions the isentropic expansion starts in
the QGP phase at a temperature of $T_0$$=$197~MeV with an equation of
state of massless gluons and $N_f$=2.3 light quarks. It evolves into a
mixed phase coinciding with hadro-chemical freeze-out at
$(T_c,\mu_B^c)$$=$$(175,232)$~MeV, followed by a hadronic evolution with
fixed hadron abundances using meson-chemical potentials. The largest
uncertainty in the dilepton yields is due to the fireball lifetime which
we fix at $\sim$7~fm/$c$ implying thermal freezeout at
$T$$\simeq$120~MeV.  The radial flow leads to moderate blue shifts of
the e.m. radiation. For the NA60-detector acceptance we use an empirical
acceptance matrix~\cite{dam06}.

In addition to thermal emission, non-thermal sources of dileptons have
to be considered: Primordial Drell-Yan annihilation is calculated as
detailed in~\cite{Abreu:2000nj}. For primordially produced $\rho$'s
which leave the hot and dense medium without thermalizing we construct a
schematic jet-quenching model.  Starting from a power law for the
initial $q_T$ spectrum we calculate the escape probability with a
``pre-hadron'' absorption cross section of
$\sigma_{\mathrm{ph}}$$=$$0.4$~mb, and a hadronic one of
$\sigma_{\mathrm{had}}$$=$$5$~mb after a $\rho$-formation time of
1~fm/$c$. At low $q_T$ we assume a scaling of the yield with the number
of participants and at high $q_T$ with the number of collisions, with a
linear transition in the range 1~GeV$<$$q_T$$<$3~GeV. Our earlier
treatment of $\rho$ decays at thermal freeze-out in terms of thermal
emission~\cite{Rapp:1999us} has been replaced by a Cooper-Frye
prescription assuming a ``sudden freeze-out'' of the entire fireball
(entailing somewhat harder $q_T$ spectra compared to thermal emission
due to an extra $\gamma$=$q_0/M$ factor).
\begin{figure}
\begin{center}
\begin{minipage}{0.32 \textwidth}
\includegraphics[width=\textwidth]{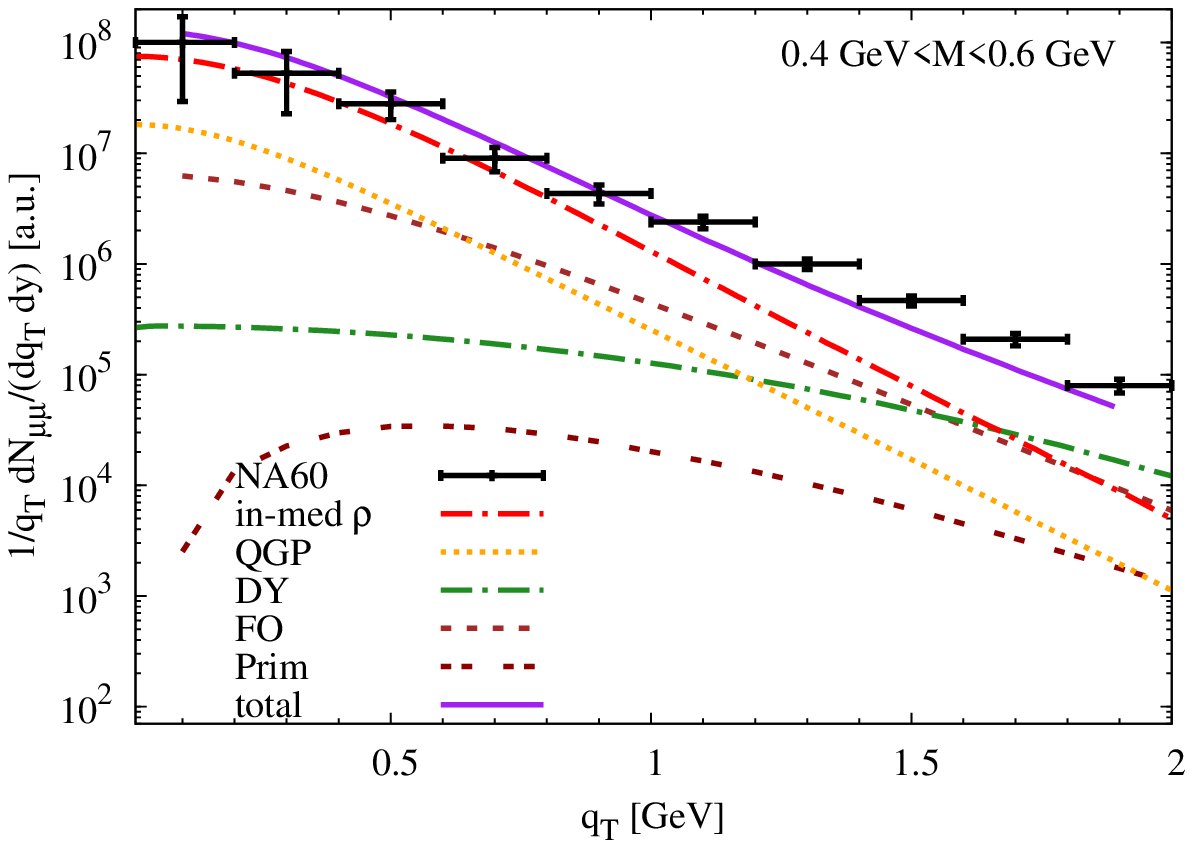}
\end{minipage}
\begin{minipage}{0.32 \textwidth}
\includegraphics[width=\textwidth]{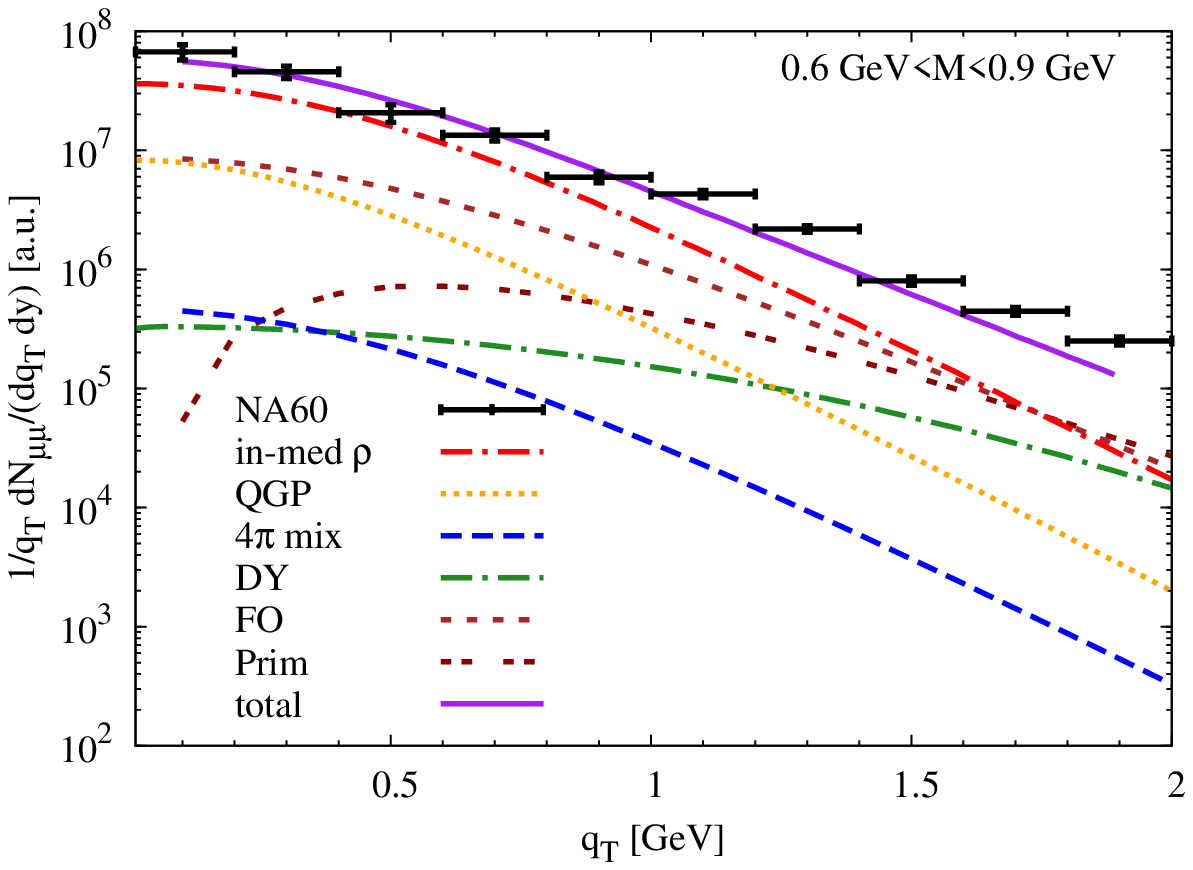}
\end{minipage}
\begin{minipage}{0.32 \textwidth}
\includegraphics[width=\textwidth]{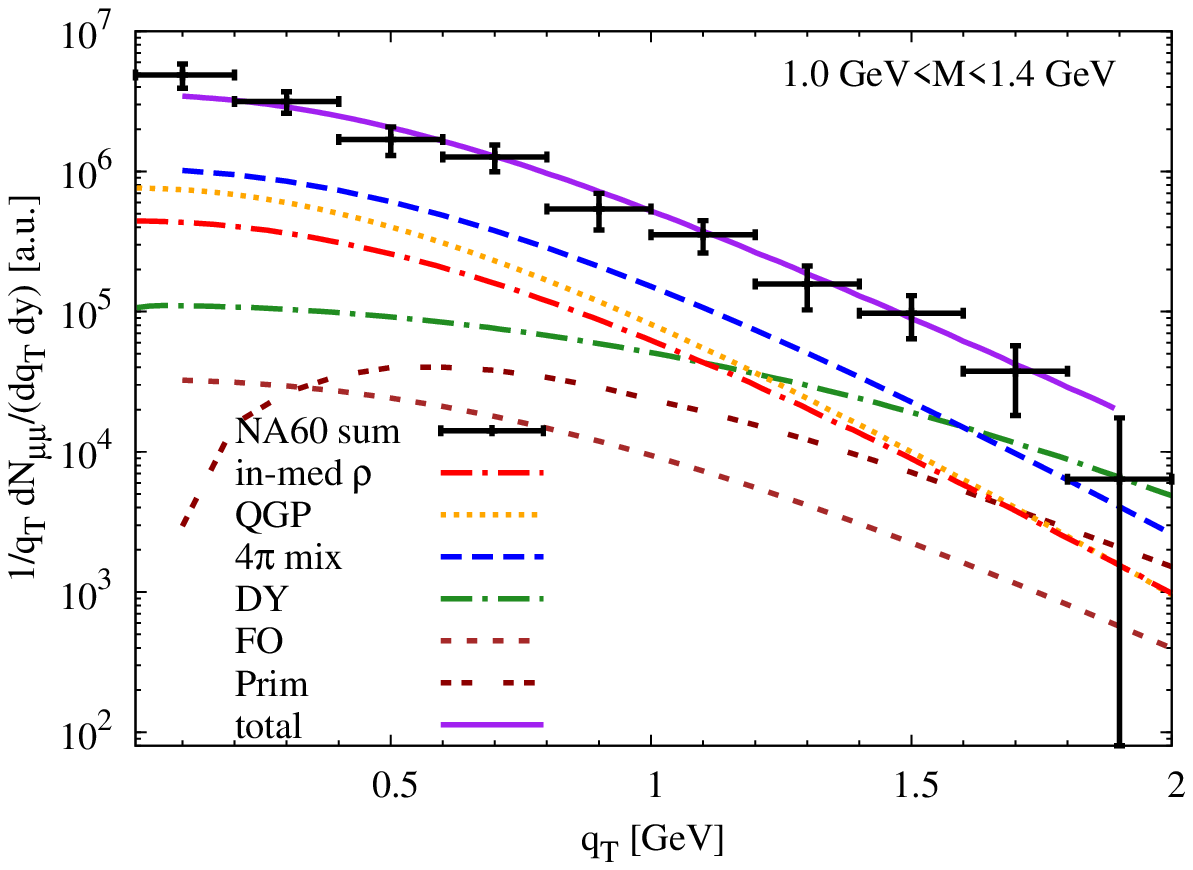}
\end{minipage}
\end{center}
\vspace*{-0.6cm}
\caption{Acceptance corrected dimuon-$q_T$ spectra in different mass
  regions (from left to right: $M=0.4$--$0.6$~GeV, $M=0.6$--$0.9$~GeV,
  $M=1$--$1.4$~GeV) in semicentral In(158~AGeV)-In
  collisions~\cite{Damjanovic:2007qm} compared to emission from QGP,
  in-medium $\rho$, $\omega$ and $\phi$ mesons, four-pion annihilation,
  primordial Drell-Yan annihilation and decays of $\rho$ mesons after
  thermal freeze-out and non-equilibrated primordial $\rho$'s.
  \label{Fig.2}}\vspace{-2mm}
\end{figure}
In the low-mass region, the NA60 invariant-mass and $q_T$ spectra are
well described within this approach,
cf.~Figs.~\ref{Fig.1}~and~\ref{Fig.2}, respectively. For the $\omega$
and $\phi$ mesons there is at present little sensitivity to their
in-medium spectral shapes. Both the $q_T$-binned mass spectra and the
$q_T$ spectra show the importance of contributions from non-equilibrated
$\rho$ mesons, i.e., those escaping the medium without (primordial
$\rho$s) and those decaying after thermal freeze-out, which both
exhibit harder $q_T$ spectra than thermal radiation. The freeze-out
contribution also has been found to be significant
in~\cite{Renk:2006qr}. In the intermediate-mass region the major sources
of dileptons are four-pion annihilation as well as correlated $D\bar{D}$
decays and radiation from the QGP.

\section{Conclusions and Outlook}
Hadronic many-body models for in-medium vector mesons provide a fair
description of dilepton emission at the SPS. High-$q_T$ sources and
centrality dependencies remain to be scrutinized. Comparisons to
upcoming RHIC data are much looked forward to.

The next goal for theory must be to implement the description of the
e.m.  current correlator within chiral models in order to establish a
direct connection to the QCD phase structure, i.e., to chiral-symmetry
restoration. Here, a promising ansatz is to constrain chiral hadronic
models by Weinberg sum rules~\cite{wein67,KS93} which relate chiral
order parameters of QCD to vector- and axial-vector current correlation
functions.

\vspace*{5mm}
\noindent \textbf{Acknowledgment.} This work has been supported by a US
National Science Foundation (NSF) CAREER Award under grant PHY-0449489.


\end{document}